# Knowledge of Songket Cloth Small Medium Enterprise Digital Transformation

**Leon A. Abdillah**[1,5,6]*, **Aisyah**[2], **Wahdyta Putri Panggabean**[3], **Sayfiyev Eldor Erkinovich**[4]
[1,2,3]Department of Information Systems, Universitas Bina Darma, Indonesia
[4] Tashkent University of Information Technologies (TUIT) Named After Muhammad al-Khwarizmi, Uzbekistan
[5] Research Fellow INTI International University (INTI IU), Malaysia
[6] Research Fellow Chung Hua University, Taiwan
[1] leon.abdillah@yahoo.com, [2] ssaishh196@gmail.com, [3] wahdytapanggabean@gmail.com
[4] eldor1075@mail.ru

**Abstract:** This article examines the knowledge of digital transformation of Small and Medium Enterprises (SMEs) that specialize in traditional handicrafts, with a specific emphasis on the Songket textile sector. The study investigates the use of digital technologies, notably blog platforms and the e-commerce site Shopee, to improve and streamline several business processes in Songket textile SMEs. The report takes a case study approach, diving into the experiences of Songket clothing enterprises that have undergone digital transformation. Key areas studied include the use of Blog platforms for brand development, marketing, and consumer involvement, as well as the Shopee E-Commerce platform for online sales and order processing. The essay seeks to give insights into the problems and possibilities faced by Songket cloth SMEs along their digital transformation journey by conducting in-depth observation, interviews, and surveys. The findings add to the scholarly discussion on the digitization of traditional industries, with practical implications for SMEs in the Songket textile sector and other handicraft areas. This study emphasizes the necessity of using digital technologies to preserve and expand traditional crafts, while also throwing light on the potential role of prominent E-Commerce platforms like Shopee in facilitating worldwide market access for such firms.

## INTRODUCTION

The digital transformation of Songket textile small and medium enterprises (SMEs) business operations Shopee can help to achieve multiple the United Nations Sustainable Development Goals (SDGs) (United Nation, 2023), including SDG 8 (Decent Work and Economic Growth), SDG 9 (Industry, Innovation, and Infrastructure), and SDG 12 (Responsible Consumption and Production) (Korucuk et al., 2022). Digital transformation may contribute to long-term economic growth and social inclusion by enabling SMEs to reach their full potential in the global market. Furthermore, digital transformation may assist to foster innovation and infrastructure development, therefore contributing to the accomplishment of SDG 9. Finally, digital transformation may contribute to responsible consumption and production by helping SMEs to embrace sustainable business practices and lessen their environmental effect. As a result, the digital transformation of Songket textile SME business operations has the potential to help achieve multiple SDGs while also encouraging sustainable development and economic growth.

Digital transformation is an important part of the digital age, and it has the potential to change the way small and medium-sized businesses (SMEs) function. The digital revolution may boost productivity, marketing initiatives, and income. According to a study on the business potential of Riau Songket woven textile firm, employing digital marketing techniques is one of nine business strategic considerations for accomplishing short-term, intermediate-term, and long-term goals (Feri et al., 2022). Another research shows that small enterprises, such as SMEs producing traditional Palembang fabric, Jumputan, and Songket, could use digital branding to enhance sales both in number and quality (Djumrianti et al., 2022). As a result, digital transformation is an important component of SME business operations, and it may help SMEs increase their competitiveness.

The Songket cloth business is an important part of the cultural history of the Melayu people in Indonesia, originated from Palembang since the time of the Sriwijaya Kingdom, and it helps to preserve the country's rich textile traditions. However, the sector has problems in maintaining its sustainability and competitiveness in the

*Leon A. Abdillah

global market, particularly in light of dynamic environmental changes and the rising limitations encountered by small and medium companies (SMEs) (Anggiani, 2016). Digital transformation has the ability to redefine how SMEs operate, creating new opportunities for development and innovation in the sector.

The development of information technology (IT) can simplify all aspects of human life such as business, education and communication by using the internet, websites and social media as examples of the application of information technology (Agustina & Abdillah, 2023). The world of trade also has the latest trend in the form of e-commerce. E-commerce or the abbreviation for electronic commerce (electronic commerce), is a business transaction that occurs on an electronic network, such as the internet. Anyone who can access a computer, has a connection to the internet, and has a way to pay for the goods or services they purchase, can participate in e-commerce (Muhammad, 2020). E-commerce or online buying and selling transactions is one application in using the internet (Yuniarti & Abdillah, 2022). The quality of e-commerce is determined by the added value provided to products or services by integrating several components from different sources such as user interaction and service quality in e-commerce itself (Adellia & Abdillah, 2020).

This research investigates the possibility of digital transformation in Songket cloth SME business operations, with an emphasis on the usage of blogs and E-Commerce Shopee platforms. In December 2023 (Table 1) there will be five top ecommerce platforms, namely: 1) Shopee, 2) Tokopedia, 3) Lazada, 4) Blibli, and 5) Bukalapak.

Table 1. Top Indonesia E-Commerce, Shopping, and Marketplace Category

| Rank | Icon | Website |
|------|------|---------|
| 1 |  | shopee.co.id |
| 2 |  | tokopedia.com |
| 3 |  | lazada.co.id |
| 4 |  | blibli.com |
| 5 |  | bukalapak.com |

Source: https://www.similarweb.com/

The research seeks to discover strategic business elements that can help achieve short-term, intermediate-term, and long-term goals such as product innovation, digital marketing techniques, and product diversification (Feri et al., 2022). This paper examines the digital transformation of Songket cloth SMEs business operations in order to contribute to the development of a fair and prosperous society, helping SMEs to reach their full potential in the global market.

**LITERATURE REVIEW**

The literature review section emphasizes the significance of digital transformation and its influence on the sustainability and competitiveness of small and medium-sized companies (SMEs) in the Songket textile sector. Several journal papers examine the issues that SMEs confront in the sector, as well as the possibility for digital transformation to overcome these challenges, namely: 1) The Commercial Potential of Riau Malay Songket Woven Cloth Enterprise (Feri et al., 2022). This study finds that the organization has made technological changes in its manufacturing process, and it emphasizes the necessity of digital marketing tactics for meeting short-term, intermediate-term, and long-term goals, 2) Efforts of Songket Craftsmen to Maintain the Sustainability of Their Business (Sadalia et al., 2019). The purpose of this study is to assess the various efforts made by cloth weaved songket craftsmen to ensure the business's survival. It emphasizes the importance of government participation and an active role in assisting craftspeople in marketing fabric songket so that they may sell songket and contribute to the business's survival, 3) Songket Handycraft Home Industries in Palembang: A Case Study (Anggiani, 2016). This report explores the issues that home sector enterprises confront as small businesses in a rapidly changing world. It underlines the significance of digital transformation in solving these difficulties and boosting SMEs' competitiveness in the global marketplace, and 4) Jumputan and Songket use a digital branding model (Djumrianti et al., 2022). This study emphasizes the paucity of research on digital branding for traditional textiles like Jumputan and Songket. It presents a digital branding plan for Jumputan and Songket textiles with the goal of improving their market position and increasing sales volume and quality.

These journal papers underline the importance of digital transformation in the Songket textile SME business, particularly in tackling the issues that SMEs confront in ensuring their sustainability and competitiveness in the global market. The papers also emphasize the effectiveness of digital marketing tactics, product innovation, and digital branding in meeting short-term, medium-term, and long-term goals.

**METHOD**

The methodologies used to investigate the digital transformation of small and medium-sized firms (SMEs) in the Songket textile industry included a variety of research tools and data gathering strategies. This section is organized into three sub-sections: 1) Research Design, 2) Data Collection Methods, and 3) Digital Transformation Stages. These various study techniques and approaches gave extensive insights into the digital transformation of SMEs in the songket textile business, offering light on the difficulties, possibilities, and prospective solutions for growth and sustainability.

**Research Design**

The research approach in this study uses a mixed methods approach. This study uses a combination of exploratory research (to better understand the present state of knowledge) and descriptive research (to examine the digital transformation of Songket textile SMEs).

**Data Collection Methods**

This research involved four data collection methods, namely: 1) Interview, 2) Survey, 3) Focus Group Discussion (FGD), and 4) Observation.

Interviews: In-depth interviews will be performed with key stakeholders to gain qualitative insights into the difficulties, accomplishments, and future plans for digital transformation. Conduct interviews with important stakeholders, including Songket cloth SMEs' owners and employees. Use open-ended questions to get qualitative information about their understanding and experiences with digital transformation.

Surveys: Create and disseminate an online questionnaire to a representative sample of Songket cloth SMEs. The survey should contain questions regarding their present understanding and use of digital technology, problems encountered, and opinions of the advantages of digital transformation.

Focus Groups: Arrange focus group meetings with many SMEs from the Songket cloth business. Encourage participants to express their perspectives, ideas, and experiences with digital transformation. This strategy can help develop ideas and provide detailed qualitative data.

Observations: Describe any field observations made during Songket workshops or on internet platforms.

**Digital Transformation Stages**

Three Stages of Digital Transformation (Bonnet, 2022), namely: 1) Modernization: This phase consists of streamlining and digitizing current processes and functions, 2) Enterprise-wide Transformation: The focus is on changing the current business, and 3) New Business Creation: The last step entails the creation of new business ventures and the discovery of new prospects.

**RESULT**

The purpose of this study was to explore the knowledge of digital transformation among Songket textile SMEs in Palembang, South Sumatra. Using a mixed-methods approach, the study provided valuable insights into these firms' awareness, comprehension, and use of digital technology.

**SME Business Modeling for Selling Songket Cloth on the Shopee E-Commerce Platform**

The Songket cloth business model (Fig. 1) in the Shopee store reflects an innovative approach in the electronic commerce ecosystem. Shopee, as a leading e-commerce platform, adopts a comprehensive business model with a focus on users and sellers. Through features such as Shopee mall, flash sale, and Shopee live, this platform facilitates active interaction between consumers and sellers. Shopee's business modeling is based on the use of advanced technology, including artificial intelligence and data analysis, to enhance the online shopping experience. Apart from that, Shopee also offers efficient logistics solutions and incentive programs for sellers, encouraging the growth of the business ecosystem within it. Overall, Shopee's business modeling reflects active involvement in connecting consumers and sellers as well as the application of technology to increase the efficiency and attractiveness of their e-commerce platform. The following is a picture of the Songket cloth business modeling at the Shopee shop and the process.

The search results shed light on the digital transformation of small and medium-sized firms (SMEs) in the Songket fabric industry, notably in locations like Jambi, Terengganu, and Sidemen Village. The articles and research studies emphasize the importance of digitization in empowering SMEs and protecting the cultural legacy related with Songket fabric. The integration of e-commerce platforms such as Shopee, web-based marketing, and the use of digital technology for product promotion, design, and online sales are all important topics stressed in literature. Furthermore, the study emphasizes the necessity of industrial clustering, design innovation, and online marketing to improve the competitiveness of SMEs in the Songket fabric business. The findings highlight the potential of digitization in pushing SMEs to continue producing items based on local culture, so contributing to the songket fabric business's sustainability and growth.

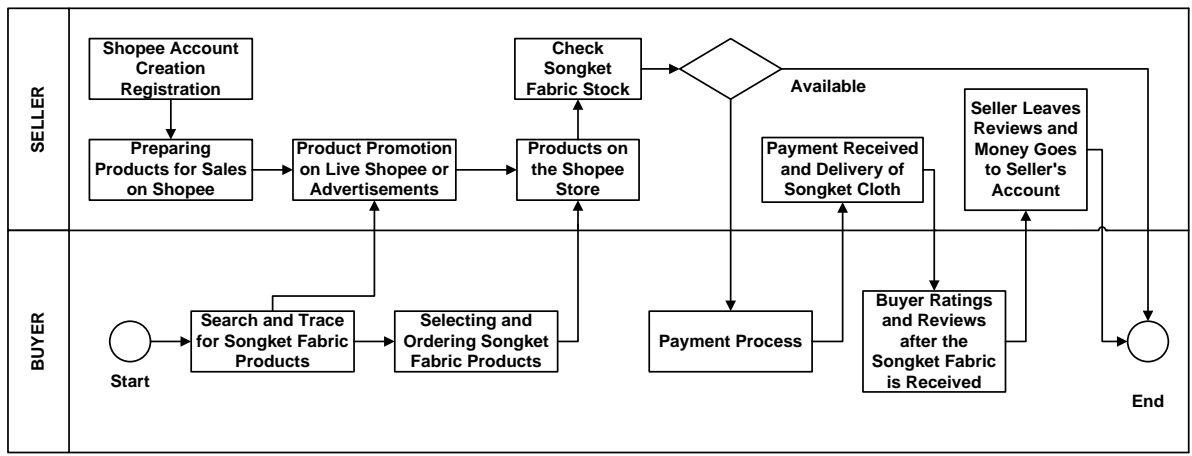

**Fig. 1** Shopee E-Commerce SME Songket Fabric Business Modeling

**SME Songket Cloth Sales Showcase on E-Commerce Shopee**

The digital transformation of small and medium-sized companies (SMEs) in the Songket cloth sector, notably in Palembang, has resulted in the emergence of home industries and SMEs that specialize in Songket manufacture and trading. Efforts to ensure the long-term viability of Songket enterprises have been recognized, with a focus on cultural heritage preservation and the necessity for government involvement in marketing and income generating. Furthermore, the integration of digital technology, social media platforms, and e-commerce has played a pioneering role in inspiring SMEs to continue producing items based on local culture, such as Songket woven fabrics. The development of Songket enterprises in different locations demonstrates the rising use of digital tactics within the sector, eventually contributing to the sustainability and expansion of the Songket fabric (Febrianti & Aprilia, 2022; Sadalia et al., 2019).

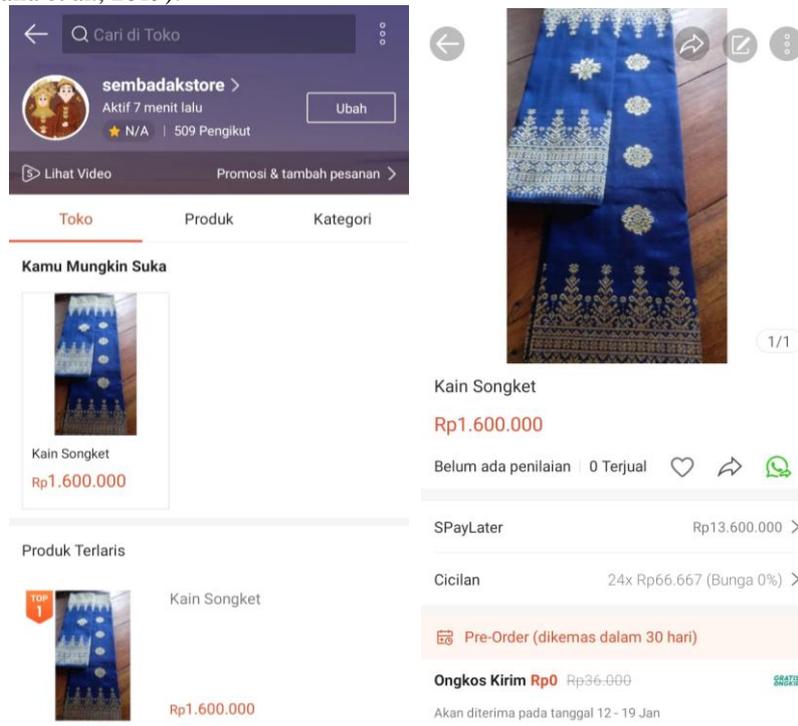

**Fig. 2** Songket Fabric products sold on Shopee

**Songket Fabric Promotion Using Blog**

One of the study's important results was that a substantial proportion of Songket cloth SMEs had yet to use digital media, such as blogs, to promote their products. However, among SMEs that have implemented digital technology, the usage of blogs has yielded good results for marketing Songket textiles.

*Leon A. Abdillah

612

Many Songket cloth SMEs have focused conventional marketing strategies such as word-of-mouth referrals and in-store displays. However, as consumer behavior becomes more digitalized, it is critical for small SMEs to discover and leverage digital channels in order to maximize their promotional efforts.

Blogs are a useful and cost-effective approach to reach a larger audience while highlighting the uniqueness of Songket textiles. By providing bright and engaging blog material, SMEs may educate potential buyers about the cultural importance, workmanship, and quality of Songket textiles, as well as showcase their current and stylish qualities.

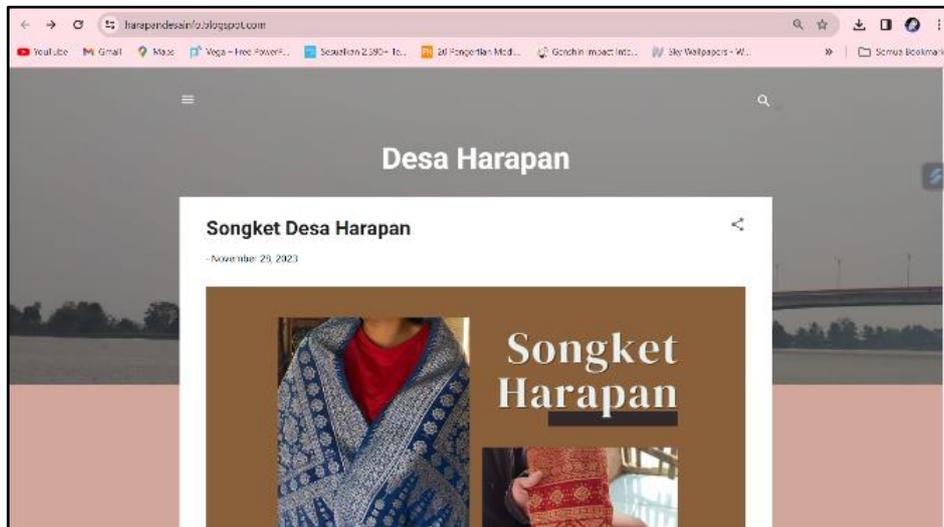

**Fig. 3** Shopee E-Commerce SME Songket Fabric Business Modeling

SMEs may utilize visually attractive photographs, videos, and educational articles to build a captivating narrative about Songket textiles, attracting both local and international buyers. Blogs also give chances for engagement and feedback, allowing prospective buyers to ask questions, seek customisation, and post testimonials, so enhancing the relationship between SMEs and their consumers.

It is vital to note that good blog marketing necessitates a mastery of search engine optimization (SEO) strategies in order to ensure that the material reaches the intended audience efficiently. Furthermore, consistent and regular blog updates are required to keep readers engaged and drive traffic to SMEs' websites or online marketplaces.

Overall, the data indicate that Songket textile SMEs would benefit substantially from adopting blogging within their digital transformation plan. By adopting this digital channel, SMEs may not only promote awareness about Songket textiles' rich legacy, but also reach a larger client base, resulting in increased sales and long-term success for their company.

**SWOT Analysis**

The SWOT analysis is an effective method for determining the strengths, weaknesses, opportunities, and threats of small and medium-sized companies (SMEs) in the Songket cloth industry. The study was used in several studies to uncover the key aspects of SMEs in the sector, including Palembang Malayu Songket Woven Cloth Enterprise (Fig. 4).

The SWOT analysis is an effective method for determining the strengths, weaknesses, opportunities, and threats of small and medium-sized companies (SMEs) in the Songket cloth industry. The study was used in several studies to uncover the key aspects of SMEs in the sector, including Palembang Malayu Songket Woven Cloth Enterprise.

Strengths (S) are the internal features and characteristics that provide a firm or project with a competitive edge: 1) Superior Product Quality: The Songket fabric offered has high quality with a unique and traditional design, which can be its own excellence and attraction compared to other Shopee products, thus providing added value for consumers, 2) Global Access: Shopee provides a platform that can be accessed globally, thus enabling the sale of Songket fabrics to reach the international market, and 3) Ease of Transactions: Digital payment facilities at Shopee make it easier for consumers to make transactions, anytime and anywhere.

Weaknesses (W) are internal issues that put a company at a competitive disadvantage: 1) Limited Stock: Songket fabrics are often produced in small quantities, which can cause limited stock for online sales, and 2) Shipping Costs: High shipping costs can be a barrier for some consumers.

Opportunities (O) are external variables that the corporation can leverage to strengthen its position: 1) Increase in Online Sales: The ever-increasing growth of e-commerce provides an opportunity to increase sales of Songket cloth, and 2) Promotional Campaign: Utilize promotional campaigns in Shopee to attract new customers.

---

*Leon A. Abdillah

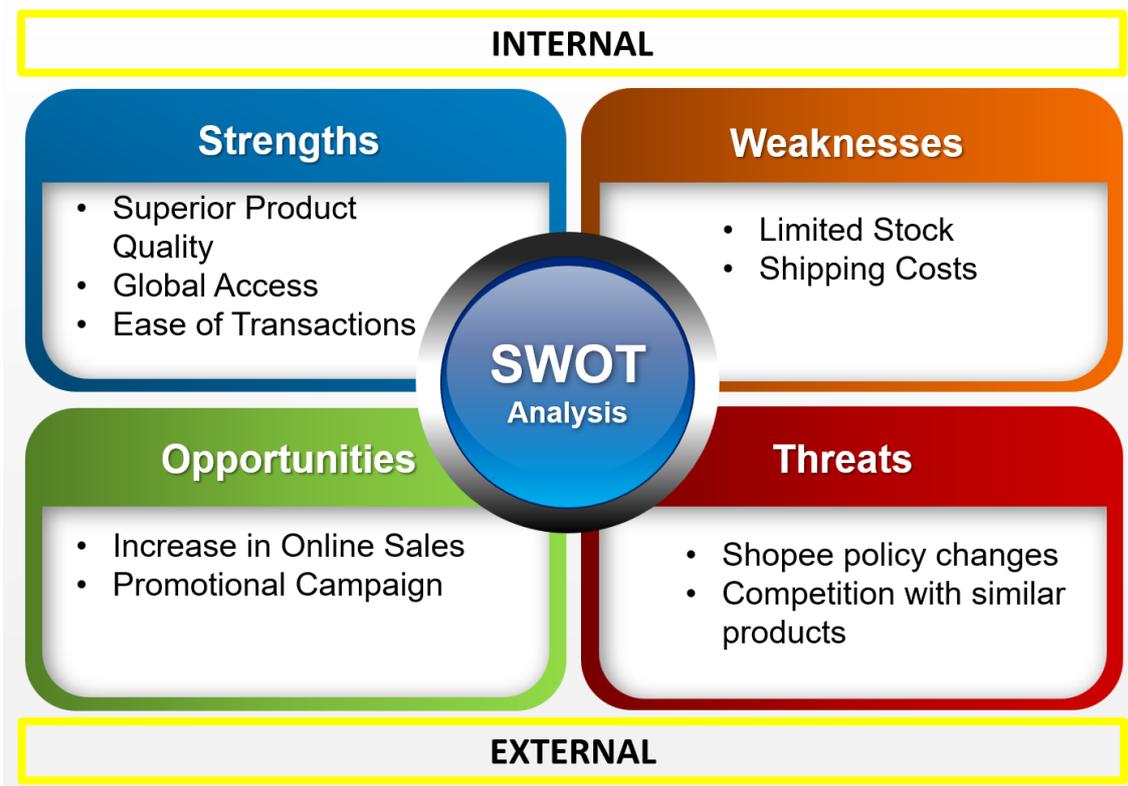

**Fig. 4** SWOT Analysis

Threats are external elements that might threaten an organization or initiative: 1) Shopee policy changes: The threat of Shopee policy or algorithm changes can affect visibility and sales, and 2) Competition with similar products: The threat of competition with other Songket moven sellers in Shopee can reduce the market share of Songket moven.

**Knowledge Management of SMEs' Digital Transformation**

The digital transformation of small and medium-sized firms (SMEs) in the Songket cloth sector has resulted in the use of numerous digital technologies and strategies to improve company operations and sustainability. Digitalization may help SMEs preserve and promote traditional crafts, such as Songket woven cloth. The integration of digital technology, social media platforms (Abdillah, 2022), and e-commerce has improved the supply chain and consumer preferences in the Songket business, indicating a growing acceptance of digital tactics.

Knowledge management (Abdillah, 2023; Rinaldi et al., 2023) is an essential component of SMEs' digital transformation in the Songket cloth sector. The growth of home industries and SMEs focused on Songket production and trade, particularly in regions such as Jambi, has highlighted the growing emphasis on empowering these businesses and increasing their competitiveness through product diversification and the incorporation of new tools and technologies to boost production capacity. The study on the business potential of Palembang Melayu Songket Woven Cloth Enterprise has emphasized the importance of innovation, digital marketing, and government support for the success of SMEs in the Songket cloth industry, particularly in expanding their market reach and international presence.

To summarize, the expertise necessary for SMEs' digital transformation in the Songket textile business comprises digital marketing, social media, e-commerce, and technology adoption. Digitalization, innovation, and government assistance have been identified as possible motivators for SMEs to retain and expand their companies in the Songket cloth sector, eventually contributing to the growth and sustainability of this traditional craft.

**DISCUSSIONS**

The digital transformation of small and medium-sized firms (SMEs) in the Songket cloth sector has been extensively researched, providing light on a variety of factors influencing the industry's growth and sustainability. Research into the innovative role of digitization in driving SMEs to sustain items based on local culture, such as

*Leon A. Abdillah

Songket woven cloth, has demonstrated the promise of digital technology in maintaining and promoting traditional crafts.

Furthermore, the business potential of Palembang Melayu Songket Woven Cloth Enterprise has been evaluated, emphasizing the importance of innovation, digital marketing, and government support in the success of SMEs in the Songket cloth industry, particularly in expanding their market reach and international presence.

The integration of digital technology, social media platforms, and e-commerce has played a vital role in boosting the supply chain and consumer preferences in the business, demonstrating the rising acceptance of digital strategies within the retail industry.

Efforts of Songket artisans to ensure the sustainability of their companies have been a primary focus, emphasising the dedication to maintaining cultural heritage and the necessity for government engagement in marketing and income creation for SMEs in the Songket cloth sector.

The growth of home industries and SMEs focused on Songket production and trade, particularly in regions such as Jambi, has highlighted the growing emphasis on empowering these businesses and increasing their competitiveness through product diversification and the incorporation of new tools and technologies to boost production capacity.

These findings highlight the potential of digitalization, innovation, and government assistance in driving SMEs to retain and expand their Songket cloth companies, therefore helping to the growth and sustainability of this traditional art.

## CONCLUSION

The conclusion that can be drawn from this service to the community is that the empowerment of Songket cloth SMEs in the Village based on e-commerce Shopee is a strategic step that illustrates a positive adaptation to the digital era. Involvement in the e-commerce ecosystem not only provides wider access, but also provides opportunities for the improvement of online business management skills. Thus, this empowerment not only increases the competitiveness of Songket cloth SMEs in the local market, but also opens up opportunities for global market exploration.

Based on the findings, the essay makes recommendations for governments, industry stakeholders, and SMEs to effectively manage the changing digital world. The findings of this study seek to contribute to the long-term growth and competitiveness of Songket Cloth SMEs in the digital era. The study emphasizes the need of embracing digital technology and investigating creative advertising tactics in Songket cloth SMEs. The promotion of Songket textiles via blogs emerged as a particularly viable method. By efficiently exploiting digital platforms, SMEs may improve their market presence, attract a larger audience, and generate long-term growth for their enterprises. Finally, the knowledge necessary for SMEs to digitally change in the Songket textile business comprises digital marketing, social media, e-commerce, and technological adaption. Future interventions should prioritize solving SMEs' knowledge gaps, offering educational assistance, and promoting the adoption of digital techniques to enable a successful digital transformation in the Songket textile business.

## ACKNOWLEDGMENT

We would like to convey our heartfelt thanks to the distinguished individuals who played critical roles in the accomplishment of this research. Special thanks go to the Kepala Desa (Village Head) and Perangkat Desa (Village Officials) from Desa Sembadak and Desa Harapan in Kecamatan Pemulutan, Kabupaten Ogan Ilir. Their continuous support, collaboration, and readiness to contribute vital ideas have significantly increased the breadth and quality of our study.